\documentclass{article}

\usepackage[nonatbib, final]{d3s3_neurips_2024}

\usepackage[utf8]{inputenc} 
\usepackage[T1]{fontenc}    
\usepackage{hyperref}       
\usepackage{url}            
\usepackage{booktabs}       
\usepackage{amsfonts}       
\usepackage{nicefrac}       
\usepackage{microtype}      
\usepackage{xcolor}         
\usepackage{graphicx}
\usepackage{subcaption}
\usepackage{amsmath}

\title{Generative Neural Reparameterization for Differentiable PDE-Constrained Optimization}

\author{%
  Archis S.~Joglekar \\
  Ergodic LLC, Seattle, WA\\
  Pasteur Labs, Brooklyn, NY\\
  Institute for Simulation Intelligence, Brooklyn, NY \\
  \texttt{archis@ergodic.io} \\
}

\begin{document}
\maketitle

\begin{abstract}
Partial-differential-equation (PDE)-constrained optimization is a well-worn technique for acquiring optimal parameters of systems governed by PDEs. However, this approach is limited to providing a single set of optimal parameters per optimization. Given a differentiable PDE solver, if the free parameters are reparameterized as the output of a neural network, that neural network can be trained to learn a map from a probability distribution to the distribution of optimal parameters. This proves useful in the case where there are many well performing local minima for the PDE. We apply this technique to train a neural network that generates optimal parameters that minimize laser-plasma instabilities relevant to laser fusion and show that the neural network generates many well performing and diverse minima.
\end{abstract}

\section{Introduction}

While Automatic Differentiation (AD) is not new \cite{griewank_evaluating_2008} and scientific programmers have exploited implementations of it for some time now, AD tooling has become much more accessible as a byproduct of the deep learning revolution in machine learning. The maturation of AD tooling has led to the paradigm of differentiable programming, where general purpose programs can be written using an AD-capable framework and then benefit from natively supporting machine learning and gradient-descent. At this point, this paradigm has been exploited by many disciplines and it would be disingenuous to attempt to enumerate them here. 

Partial-differential-equation (PDE) constrained optimization is pertinent to many scientific and engineering disciplines \cite{biegler2003large, de2015numerical}. A key component of PDE-constrained optimization is the method by which the gradient is computed. When a numerical program that solves PDEs is written using an AD-capable framework, it immediately lends itself to PDE constrained optimization. 

In this work, we will show that when PDE constrained optimization is performed using an AD-capable framework i.e. when the PDE solver is made differentiable, the benefit is not only access to a fast and accurate gradient with respect to the optimization parameters, but also the ability to learn distributions of optimal parameters via highly-parameterized neural networks. When the neural network is made `generative', i.e. when it maps from a random vector, it can help capture degeneracy in PDE-constrained optimization problems with many local minima.

\section{Generative PDE-constrained optimization}
\begin{figure}[htbp]
\begin{center}
\includegraphics[width=0.8\textwidth]{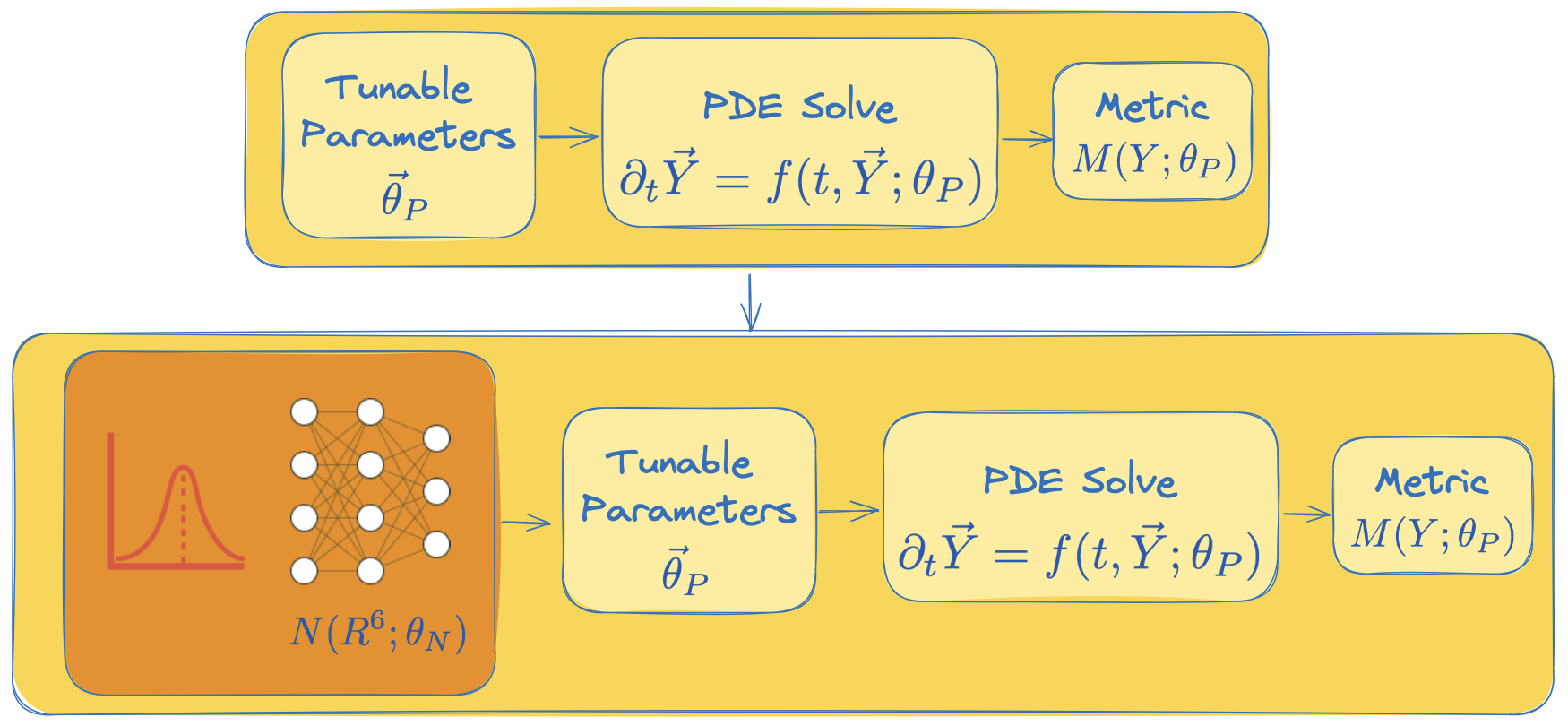}
\end{center}


\caption{(a) The usual PDE constrained optimization problem where the parameters, $\theta_P$, are optimized to minimize the metric value, $M$. The required gradient is $\partial M / \partial {\theta_P}$. (b) Generative neural PDE constrained optimization comprises a neural network that maps a random variable to the optimal parameter set $\theta_P$. The relevant gradient is now $\partial M / \partial {\theta_N}$.}
\label{fig:gnr}
\end{figure}

In this section, we explain our method, how it is an extension of previous work, and the advantages it offers in comparison to direct PDE constrained optimization. In the discussion that follows, $Y(T, \vec{x})$ is the solution to a PDE parameterized by parameter set, $\theta_P$. $\mathcal{M}$ is the function that computes the scalar quantity that is to be minimized. The minimization strategy is first-order gradient descent.

\subsection{Direct method}
Here, direct optimization refers to solving the optimization problem posed by 
\begin{align}
    \min_{\theta_P} ~~~ &\mathcal{L} \left[Y(T, \vec{x}; \theta_P)\right] \label{eq:directopt} \\
    s.t. ~~~ &\partial_t Y = f(t, \vec{x}, Y(t, \vec{x}); \theta_P) \label{eq:pdeconst}.
\end{align}
This is a typical PDE constrained optimization problem. While effective, performing gradient descent to solve eq. \ref{eq:directopt} to obtain a distribution of optimal parameters can require many iterations and suffer from the curse of dimensionality because one optimization produces one set of optimal parameters. 

\subsection{Neural Reparameterization (NR)}
In ref. \cite{hoyer_neural_2019}, it was proposed that the parameter set be reparameterized using a neural network. For that particular application, the reparameterization enabled the optimization algorithm to uncover better performing minima. Formally, they solved the optimization problem posed by
\begin{align}
    \min_{\vec{g}, \theta_N} ~~~ &\mathcal{L} \left[Y(T, \vec{x}; N(\vec{g}; \theta_N))\right]  \\
    s.t. ~~~ &\partial_t Y = f(t, \vec{x}, Y(t, \vec{x}); N(g; \theta_N)) \label{eq:nr},
\end{align}
where $\theta_P = N(\vec{g}; \theta_N)$ is the neural reparameterization (NR) of $\theta_P$. Crucially, in that work, the input vector, $\vec{g}$, to the neural network is a learned quantity in addition to the weights of the neural network $\theta_N$. Like with Direct method above, NR is also unable to obtain multiple local minima without suffering from the same problems.

\subsection{Generative Neural Reparameterization (GNR)}
Here, we propose that $\vec{g}$ is no longer learned, but is instead a random variable. This results in the neural network being responsible for learning a mapping from a random variable to an optimal set of parameters. Figure \ref{fig:gnr} provides a conceptual illustration of this method. Formally, the problem is stated by 
\begin{align}
    \min_{\theta_N} ~~~ &\mathcal{L} \left[Y(T, \vec{x}; N({G}; \theta_N))\right]  \\
    s.t. ~~~ &\partial_t Y = f(t, \vec{x}, Y(t, \vec{x}); N(G; \theta_N)) \label{eq:gnr},
\end{align}
where $G$ is now a random vector from an $D$-dimensional unit normal vector where $D$ is another hyperparameter for the neural network. Here, we choose $D=6$. We find the training process to be relatively insensitive to this parameter within $4 < D < 16$. This addition to the NR method enables learning distributions of optimal parameters. The neural network is responsible for transforming the unit normal probability distribution to the probability distribution of optimal spectra. 

If the optimization problem has many satisfactory local minima and many parameters where gradient-descent is necessary for the problem to be tractable, each instantiation of the direct method and NR will converge to a single minimum from the set of satisfactory minima. On the other hand, GNR learns distributions of optimal parameters. In addition to obtaining a generative function for inverse design, learning a distribution of optimal solutions can help with downstream analyses of the behavior of the solution at the optimal point.


In the following section, we apply the GNR method to a PDE-constrained minimization problem where local minima abound.

\section{Application - Laser Plasma Instability Modeling}
\subsection{Background}
Laser-driven inertial fusion has demonstrated promising recent results in the laboratory \cite{indirect_drive_icf_collaboration_lawson_2022, zylstra_burning_2022} and efforts are underway to scale these experiments into those that can deliver enough power to serve as part of a fusion power plant. One of the persistent challenges with scaling the experiments and delivering large amounts of laser energy to the fusion target is the presence of laser plasma interactions. Plasmas are electromagnetic media which can and do interact with lasers. While laser plasma interactions have been used to make implosions more symmetric by leveraging plasma's ability to transfer energy between laser beams \cite{dewald_early-time_2013}, they are typically detrimental in other aspects of the laser plasma physics that occurs during such experiments. For this reason, it is important to model the behavior of, and minimize the impact of, these laser plasma instabilities that plague laser-driven inertial fusion.

Recent advances in laser technology have enabled the development of broadband high energy lasers. Adding bandwidth, i.e. where the laser is not monochromatic but has high energy and intensity light at many wavelengths, helps mitigate laser-plasma instabilities \cite{follett_suppressing_2018, follett_thresholds_2019, follett_thresholds_2021}. For this reason, future laser fusion experiments and facilities will be designed with bandwidth. It will be important to determine exactly what kind of bandwidth spectra best serve the purpose of transferring laser energy to the target through the plasma. To do so, we will need bandwidth that helps minimize the interaction between the laser and the plasma. In the following section, we detail the PDE and the ensuing laser plasma instability minimization problem. 

\subsection{Equations}
We model one such laser-plasma instability, called Two Plasmon Decay, using the PDE that describes electron plasma waves, also given in ref. \cite{follett_thresholds_2019}, given by
\begin{align}
    \nabla \cdot &\left[ i \left(\frac{\partial}{\partial t} + \nu_e ~ {\circ} \right) + \frac{3 v_{te}^2}{2 \omega_{p0}} \nabla^2 + \frac{\omega_{p0}}{2}\left(1-\frac{n_b(x)}{n_0}\right) \right] \textbf{E}_h = S_{\text{TPD}}(E_0) + S_h
\end{align}
where the source term, $S_\text{TPD}$, contains $E_0$, the laser field that can be optimized. The equations are described in more detail in ref. \cite{follett_thresholds_2019} and the appendix. It is the laser field that contains the free parameters that can be optimized to minimize laser-plasma instability activity. The laser field is a sum of $N_c$ complex fields and is given by
\begin{align}
    E_0(t, x) = \sum_j^{N_c} {\color{red}{a_j}} \exp( i k_j(x) x - i \omega_j t + {\color{red}{\phi_j}}).
\end{align}

Formally, the optimization problem of interest is given by
\begin{align}
    \min_{a_j, \phi_j} &\int_{15\text{ps}}^{20\text{ps}} \int \int d_t|\textbf{E}_{h}(t, x, y; a_j, \phi_j, N_c=8)|^2 ~ dx ~ dy ~ dt \label{eq:do},
\end{align}
where we are interested in minimizing the growth rate of the instability. 

Performing GNR on eq. \ref{eq:do} gives
\begin{align}
    \min_{\theta_N}~~~~~ &\int_{15\text{ps}}^{20\text{ps}} \int \int d_t|\textbf{E}_{h}(t, x, y; a_j, \phi_j, N_c=8)|^2 ~ dx ~ dy ~ dt\\
    s.t.~~~~~ &a_j = f(N(G); \theta_N) \text{ and } \phi_j = g(N(G); \theta_N). \label{eq:gnrlpi}
\end{align}

\subsection{Results}
\begin{figure}[h]
    \centering
    \includegraphics[width=0.9\linewidth]{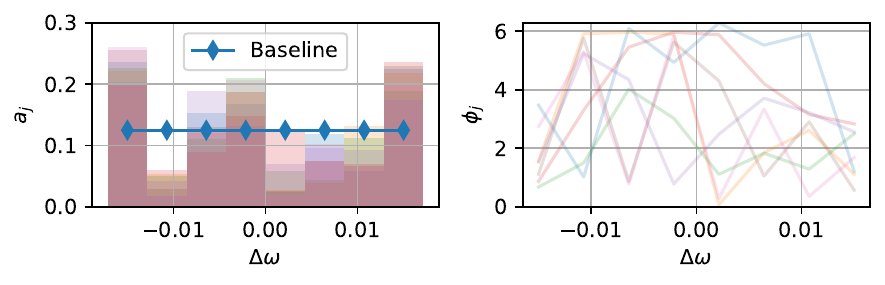}
    \caption{Example spectra from a trained GNR model}
    \label{fig:spectra}
\end{figure}

In this section, we solve eq. \ref{eq:gnrlpi}\footnote{Training and solver details are given in the Appendix} and show comparisons of the results of the optimization for $N_c = 8$ using GNR to the baseline case of a uniform amplitude, $a_j = 0.125$, spectrum with random phase $\phi_j = U[0, 2\pi]$, as in ref. \cite{follett_thresholds_2019}. An example amplitude and phase pair from the uniform case, along with six instantiations of the generated spectra, are shown in fig. \ref{fig:spectra}.

\begin{figure}[h]
    \centering
    \includegraphics[width=0.5\linewidth]{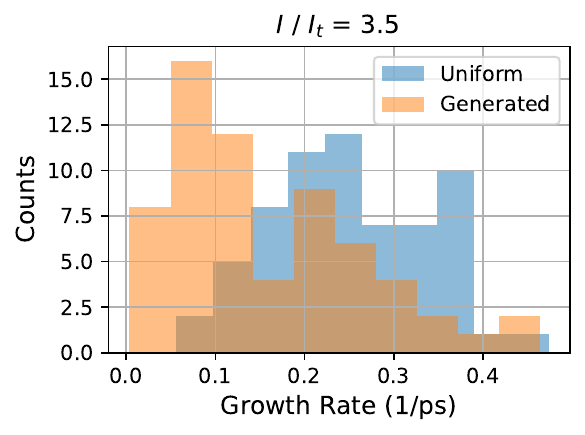}
    \caption{A comparison of the distribution of growth rates from simulations using the baseline uniform-random spectra and the spectra acquired using the GNR method shows a clear improvement over baseline.}
    \label{fig:growthrates}
\end{figure}

Figure \ref{fig:growthrates} shows the comparison of growth rates of the absolute instability from simulations with generated spectra and from simulations using the baseline. 
The figure indicates that the generated distribution generally performs better than the baseline case in task of minimizing the growth rate. The analysis of the results and the underlying physics is beyond the scope of this workshop paper. The aim here is to demonstrate that the GNR method can capture many well performing local minima in PDE-constrained optimization. 

\section{Conclusions}
PDE-constrained optimization is a common technique that can benefit from Automatic Differentiation and Differentiable Programming. In addition to the fast calculation of gradients, the ability to train neural networks in-line opens many possibilities. 
We highlight one such possibility here where a neural network can be trained to map from a random vector to a set of optimal parameters. This enables machine learning distributions of optimal parameters in the context of PDE-constrained optimization with many desirable local minima.

\begin{ack}
We acknowledge useful discussions with Dr. Peter Sharpe and Dr. Aidan Crilly. 

This material is based upon work supported by the Department of Energy Office of Fusion Energy under Award Number(s) DE-SC0024863. This research used resources of the National Energy Research Scientific Computing Center, a DOE Office of Science User Facility supported by the Office of Science of the U.S. Department of Energy under Contract No. DE-AC02-05CH11231 using NERSC award FES-ERCAP0026741. This work also received support from Pasteur Labs and ISI.

The authors declare no competing interests. 
\end{ack}

\bibliography{bib} 
\bibliographystyle{ieeetr}

\appendix
\newpage

\section{Solver Details}

This solver is available as an option within \textsc{ADEPT} \cite{joglekar2023adept}, an open-source differentiable simulation tool for plasma physics \footnote{https://www.github.com/ergodicio/adept}. 
\subsection{Equations}
The PDE for Two Plasmon Decay \cite{follett_thresholds_2019} is given by
\begin{align}
    \nabla \cdot &\left[ i \left(\frac{\partial}{\partial t} + \nu_e ~ {\circ} \right) + \frac{3 v_{te}^2}{2 \omega_{p0}} \nabla^2 + \frac{\omega_{p0}}{2}\left(1-\frac{n_b(x)}{n_0}\right) \right] \textbf{E}_h = S_{\text{TPD}}(E_0) + S_h \\
    S_{\text{TPD}} &\equiv \frac{e}{8 \omega_{p0} m_e} \frac{n_b(x)}{n_0} \nabla \cdot [\nabla (\textbf{E}_0 \cdot \textbf{E}_h^*) - \textbf{E}_0 \nabla\cdot \textbf{E}_h^*] e^{-i (\omega_0 - 2 \omega_{p0})t}
\end{align}
where the first term represents oscillations in the electron plasma, the second represents Landau damping, the third term represents the effect of a plasma density gradient, and the final term is the driving term that contains $E_0$, the laser field. It is the laser field that contains the free parameters that can be optimized to minimize laser-plasma instability activity. The laser field is given by
\begin{align}
    E_0(t, x) = \sum_j^{N_c} a_j \exp( i k_j(x) x - i \omega_j t + \phi_j)
\end{align}

\subsection{Parameters}
The simulation parameters are given in table \ref{tab:sim}.
\begin{table}[h]
    \centering
    \begin{tabular}{|| c|c || c | c ||}
        $\Delta x, \Delta y$ &  20 nm &
        $\Delta t$ & 2 fs \\
        $L_x$ & 140 $\mu$m &
        $L_y$ & 16 $\mu$m \\
        $t_{\text{max}}$ & 20 ps &
        $L_n$ & 366 $\mu$m \\
        $T_e$ & 3.92 keV & $I_0$ & $7.53 \times 10^{14}~\text{W/cm}^2$
    \end{tabular}
    \caption{Simulation Parameters, Geometry, and Discretization}
    \label{tab:sim}
\end{table}

\section{Training}
\subsection{Neural Network Details}
Two separate neural networks were trained, one each for amplitude and phase. Their parameters are given in table \ref{tab:nn}.

\begin{table}[h]
    \centering
    \begin{tabular}{|| c|c || c | c ||}
        Input & N$(0, 1)^6$ & Optimizer & ADAM \\
        Width & 128 & Learning Rate & 0.001 \\
        Depth & 3 & Activation Function & $\tanh$ \\
        Output & 8 
    \end{tabular}
    \caption{Training details}
    \label{tab:nn}
\end{table}

It is important to note that the output of the neural network responsible for computing the amplitude spectrum was exponentiated. This was done to encourage higher contrast ratios between values. The output of the neural network was constrained to $[-3, 3]$ and then exponentiated and normalized to maintain the desired total amplitude.

\end{document}